\documentclass[12pt]{iopart}

\usepackage{epsfig}
\usepackage{xcolor}
\usepackage{graphicx}
\usepackage{dcolumn}
\usepackage{bm}
\usepackage[english]{babel}
\usepackage{iopams}
\usepackage{epstopdf}

\begin{document}

\title[Viscous Properties of Nickel-Containing Binary Metal Melts]{Viscous Properties of Nickel-Containing Binary Metal Melts}

\author{R M Khusnutdinoff$^{1,2}$, R R Khairullina$^{1}$, A L Beltyukov$^{1,2}$, V I Lad'yanov$^{2}$ and A V Mokshin$^{1,2}$}
\address{$^1$Department of Computational Physics, Institute of
Physics, Kazan Federal University, 420008 Kazan, Russia}
\address{$^2$Udmurt Federal Research Center, Ural Branch of Russian Academy of Sciences, 426068 Izhevsk, Russia}

\ead{anatolii.mokshin@mail.ru}

\date{\today}

\begin{abstract}
The paper presents the results of molecular dynamics study of the viscosity of nickel-containing binary metal melts for a wide range of temperatures, including the region of the equilibrium liquid phase and supercooled melt. It is shown that the temperature dependencies of the viscosity of binary metal melts are described by the Kelton's quasi-universal model. Based on the analysis of the viscosity coefficient of the binary melt composition within the framework of the Rosenfeld's scale transfor\-mations, it has been established that to correctly describe the viscosity of binary/multi\-component metal melts within the framework of entropy models, it is necessary to use a more complex representation of the excess entropy $S_{ex}$ than in the approximation of pair correlation entropy $S_2$.
\end{abstract}


\vspace{1.0cm} \noindent{\it Keywords\/}: binary metallic alloys,
liquids, entropy, viscosity, molecular dynamics simulations

\maketitle

\section{Introduction}
Viscosity is one of the most important characteristics that determines relaxation features, thermophysical and transport properties of a substance; has a high sensitivity to structural transformations and phase transitions, and also plays an important role in the kinetics of chemical reactions~\cite{Trachenko2020}.
At the same time, the temperature dependence of the viscosity determines the so-called glass-forming ability of the system~\cite{Bellissard2018,Khusnutdinoff2020a}.
In general, the viscosity of a liquid varies with its temperature and composition and can be measured experimentally using viscometry techniques \cite{Brooks2005}, or calculated by classical/quantum molecular dynamics simulations~\cite{Koishi1996}, or obtained by means of semi-empiric or microscopic theoretical models \cite{Bellissard2018,Desgranges2008,Tlili2017}.
Indirect experimental techniques, such as inelastic scattering of neutrons, X-rays, and Brillouin light scattering, are characterized by significant inaccuracies in determining the transport coefficients (diffusion, viscosity) \cite{Scopigno2005}. At the same time, the determination of viscosity using viscometry (capillary viscometry method, torsional vibration method, ultrasonic method, etc.) is associated with significant difficulties, due primarily to low sensitivity and imperfection of experimental techniques \cite{Viswanath2007,Cheng2014}. Another alternative in finding the viscosity is the methods of classical and quantum-mechanical (\textit{first-principle}) molecular dynamics simulations, which are characterized by a number of serious limitations: first -- the accuracy and predictive ability of the interatomic interaction potentials; second -- presence of approximations in the exchange-correlation potential and the limited (short) time scales of simulations \cite{Marx}.

One of the key problems in classical molecular dynamics simulations of metallic systems is the choice of the interatomic interaction potential \cite{March1990}. Thus, for example, pair potentials incorrectly take into account the peculiarities of metallic bonds, and therefore such potentials are rarely used in the study of the properties of metallic systems. Further, recently we have shown that in the case of liquid lithium near the melting temperature, the spherical pseudopotential provides a better agreement with experimental data on elastic and inelastic X-ray scattering as compared to the known many-particle potentials of the EAM-type \cite{Khusnutdinoff2018}. At the same time, semi-empirical many-atom potentials based on the embedded atom method (EAM) and its modification (MEAM) are the most successful for describing the structural features and dynamic properties of the polyvalent metallic systems \cite{Daw1993}. The limitations of a EAM-potential are well-known: it works properly for purely metallic systems with no directional bondings; it does not treat covalency or significant charge transfer; and it does not handle Fermi-surface effects.

Despite the fact that first-principle molecular dynamics methods can overcome all these limitations, they have also some limitations.
The most modern quantum-mechanical molecular dynamics methods are based on the density functional theory \cite{Marx}. In this approach, the primary equations are cast in terms of the electron density rather than on the wave functions. However, although the density functional theory is well-developed, it remains the case that for certain components (particularly the exchange and correlation terms) exact functionals are not available. So, for example, to find the exchange-correlation energy, such approximations are used as ``local density approximation'' (LDA), ``generalized gradient approximation'' (GGA) and others \cite{Marx}.

There has been no comprehensive microscopic theory on the viscosity of liquids so far because of their structural complexity.
Therefore, many calculations were performed to predict the viscosity of multicomponent alloys using thermodynamic viscosity models \cite{Gao2019}. However, a molten alloy can not be considered as an ideal mixture, and there are often discrepancies between predictions and experiments \cite{Pasturel2015,Mokshin2020,Khusnutdinoff2016a}. On the other hand, it is also difficult to define the thermodynamic parameters within the thermodynamic viscosity models for multicomponent melt systems~\cite{Shi2019}.
Thus, the refinement of absolute values and the development of universal models of viscosity is still one of the important problems of modern thermal physics and condensed matter physics \cite{Khusnutdinoff2020}.

Nickel-containing binary metal melts were previously studied by experimental, theoretical and numerical methods using methods of classical and first-principle molecular dynamics simulations. In Ref.~\cite{Brillo2006} using X-ray diffraction, the structural features of binary $Al-Cu$ and $Al-Ni$ metal melts were investigated. The experimental results of the viscosity of binary $Al-Ni$ and $Fe-Ni$ metal melts are reported in Refs.~\cite{Kehr2008, Petrushevsky1971, Sato2005, Krieger1971, Baum1979}. A theoretical analysis of the viscosity of aluminium-based binary alloys using semi-empirical models is presented in Refs.~\cite{Gasior2014,Chen2014}. By using \textit{ab-initio} molecular dynamics simulations the transport properties and the validity of the Stokes-Einstein relation in Al-rich liquid alloys have been studied in Ref.~\cite{Jakse2016}.

The purpose of this study is to determine the absolute values of viscosity for aluminum-nickel and iron-nickel melts, as well as to check the applicability of various quasi-universal models to describe the viscosity of binary metal melts.

\section{Experimental Method\label{exper}}
Binary $Al-Ni$ alloys were prepared by melting highly pure aluminum and the $Al_{100-x}Ni_x$ (where $x=1$ at.\% and $15$ at.\%) alloys in a viscometer furnace in an atmosphere of high-purity helium at a temperature of $T=1373$~K and isothermal exposure for at least $1$~hour. When smelting alloys with nickel content from $1$ to $9$ at.\% and with nickel content less than $1$~at.\% the alloy of $Al_{85}Ni_{15}$ and $Al_{99}Ni_1$ was used respectively. The ligatures were obtained by melting metals in a resistance furnace at a residual pressure of $10^{-2}$~Pa and a temperature of $T=1943$~K for $30$~min. The initial components were highly pure aluminum ($99.999$ wt.\% $Al$) and electrolytic nickel ($99.5$ wt.\% Ni). The nickel content in the alloys was determined by atomic emission spectroscopy by means of a \textit{SPECTROFlame Modula D} spectrometer. The kinematic viscosity of the melts was measured on an automated installation by the torsional vibration method \cite{Beltyukov2008, Shvidkovsky}. The measurements were carried out in a protective atmosphere of purified helium. Cylindrical cups made of $Al_2O_3$ with an inner diameter of $17$~mm and a height of $40$~mm were used as crucibles. A lid was placed in the crucible over the sample. The lids were made from $Al_2O_3$ cups with a height of $12$~mm and an outer diameter $0.4-0.6$~mm smaller than the inner diameter of the crucible. The design of the crucible with a lid is given in Ref.~\cite{Beltyukov2019}. The lid can move along the vertical axis of the crucible, compensating for the changes of the sample volume. When performing torsional vibrations, the lid moves together with the crucible, creating an additional end surface of friction with the melt.
The crucibles and lids were preliminarily annealed in a vacuum furnace at a residual pressure of $10^{-2}$~Pa at a temperature of $T=1923$~K and isothermal exposure for $1$~hour. The use of a crucible with a lid during viscosity measurements  makes it possible to exclude the influence of film effects and wetting phenomena on the measurement results \cite{Khusnutdinoff2016}. Before measurements, all samples were remelted at a temperature of $1473$~K in a viscometer oven, followed by cooling to room temperature. Temperature dependences of the viscosity were obtained in the modes of heating from the liquidus temperature of the alloy to $1473$~K and subsequent cooling until the beginning of crystallization of the melt. The melt was isothermally exposed for 15 min at each temperature before the start of measurements. The values of the kinematic viscosity and the error in its determination were calculated according to the methods described in Refs.~\cite{Khusnutdinoff2018a,Khusnutdinoff2018b}. The general relative error at measuring the viscosity did not exceed $4$ \%, while the error of a single experiment does not exceed $2$ \%.

\section{Details of Simulation and Numerical Calculation\label{comp}}
\begin{figure*}
	\begin{center}
		\includegraphics[height=7.7cm, angle=0]{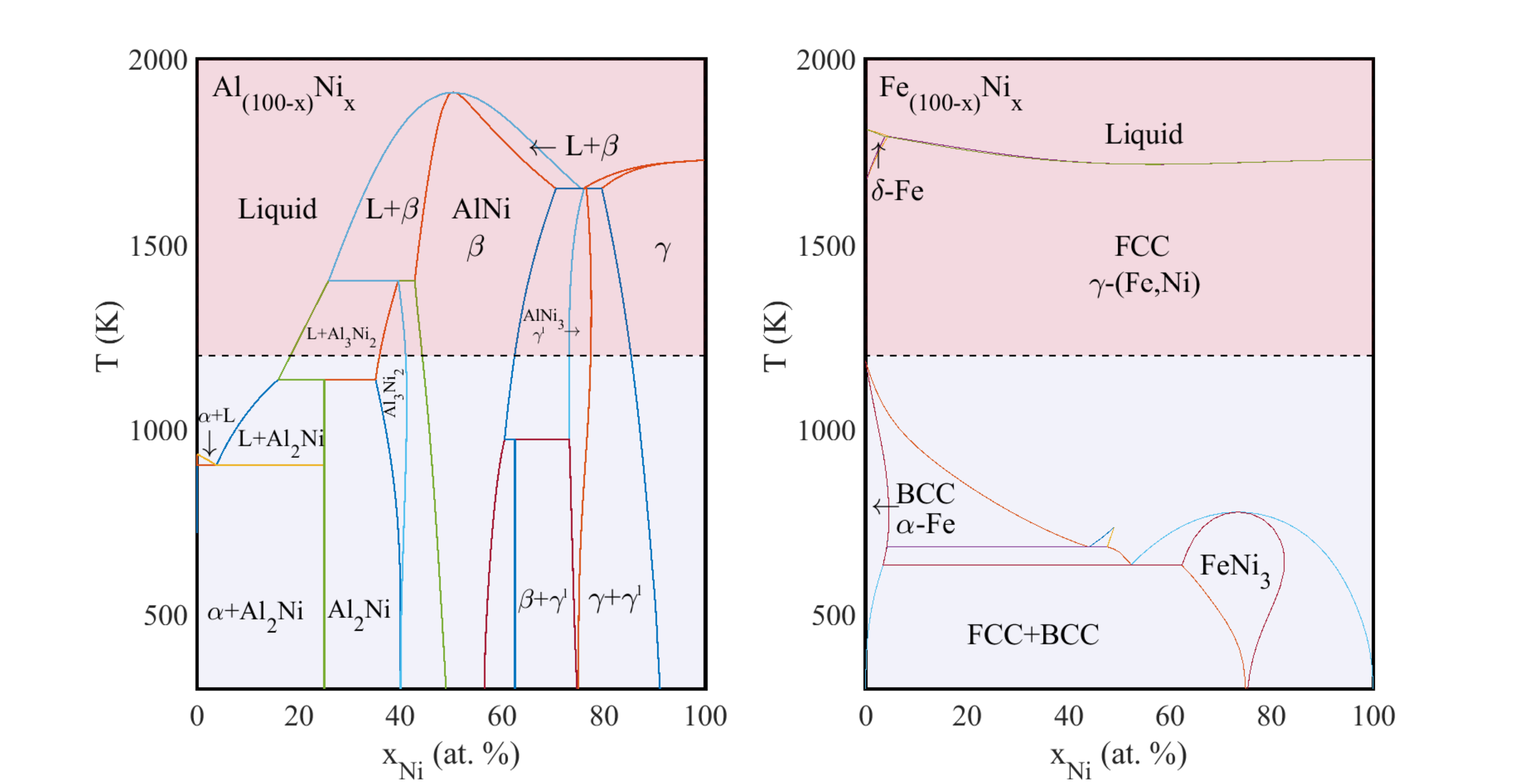}
		\caption{(Color online) Phase diagrams for the aluminum-nickel $Al_{(100-x)}Ni_x$ system (left column) and the iron-nickel $Fe_{(100-x)}Ni_x$ system (right column) \cite{Massalski1986}. The investigated region of the phase diagram in this study is highlighted in rose color.} \label{Fig_PhD}
	\end{center}
\end{figure*}
The molecular dynamics simulation of nickel-containing metal melts $Al_{(100-x)}Ni_x$ and $Fe_{(100-x)}Ni_x$ was carried out in an NpT-ensemble at a pressure of $p=1.0$~bar for the temperature range $T=[1200: 2000]$~K, which covers the region of the equilibrium liquid phase and the region of the supercooled state (see the phase diagram in Fig.~\ref{Fig_PhD}). All simulations were performed using the LAMMPS package~\cite{Plimpton1995}.
The systems under study consisted of $N=32000$ atoms allocated in a cubic cell with periodic boundary conditions. Interactions between particles were carried out using the EAM-type potentials \cite{Mishin2004} and \cite{Bonny2009a,Bonny2009b}, respectively.
The embedded-atom method (EAM) represents the potential energy $E$ of the systems in the form
\begin{equation}
E=\frac{1}{2}\sum_{i,j}\Psi_{ij}(r_{ij})+\sum_iF_i(\overline{\rho}_i).
\end{equation}
Here, $\Psi_{ij}(r_{ij})$ is the pair interaction energy between atoms $i$ and $j$ separated by a distance $r_{ij}$, $F_i$ is the embedding energy of atom $i$ and $\overline{\rho}_i$ is the ``effective'' electron density induced by all surrounding atoms $j$ at the location of atom $i$. The ``effective'' electron density is given by
\begin{equation}
\overline{\rho}_i=\sum_{i\neq j}\rho_j(r_{ij}),
\end{equation}
where $\rho_j(r)$ is the electron-density function assigned to atom $j$. The pair interaction, electron-density and embedding functions depend on the chemical sorts of atoms. A detailed description and parameters of EAM-potentials for binary metallic melts $Al_{(100-x)}Ni_x$ and $Fe_{(100-x)}Ni_x$ can be found in Refs.~\cite{Mishin2004,Bonny2009a,Bonny2009b}.

Binary melts were obtained by rapid cooling of systems from a high-temperature equilibrium state with $T=2000$~K. The cooling rate of the systems was $\gamma=1.0$~K/ps. The equations of motion of atoms were integrated using the velocity Verlet algorithm with a time step of $1.0$~fs. To bring the systems to the state of thermodynamic equilibrium, the program performed $1.5\cdot 10^7$ time steps and $2\cdot 10^8 $ steps to calculate the time correlation functions.

\section{Results}
\begin{figure*}
\begin{center}
\includegraphics[height=11.0cm, angle=0]{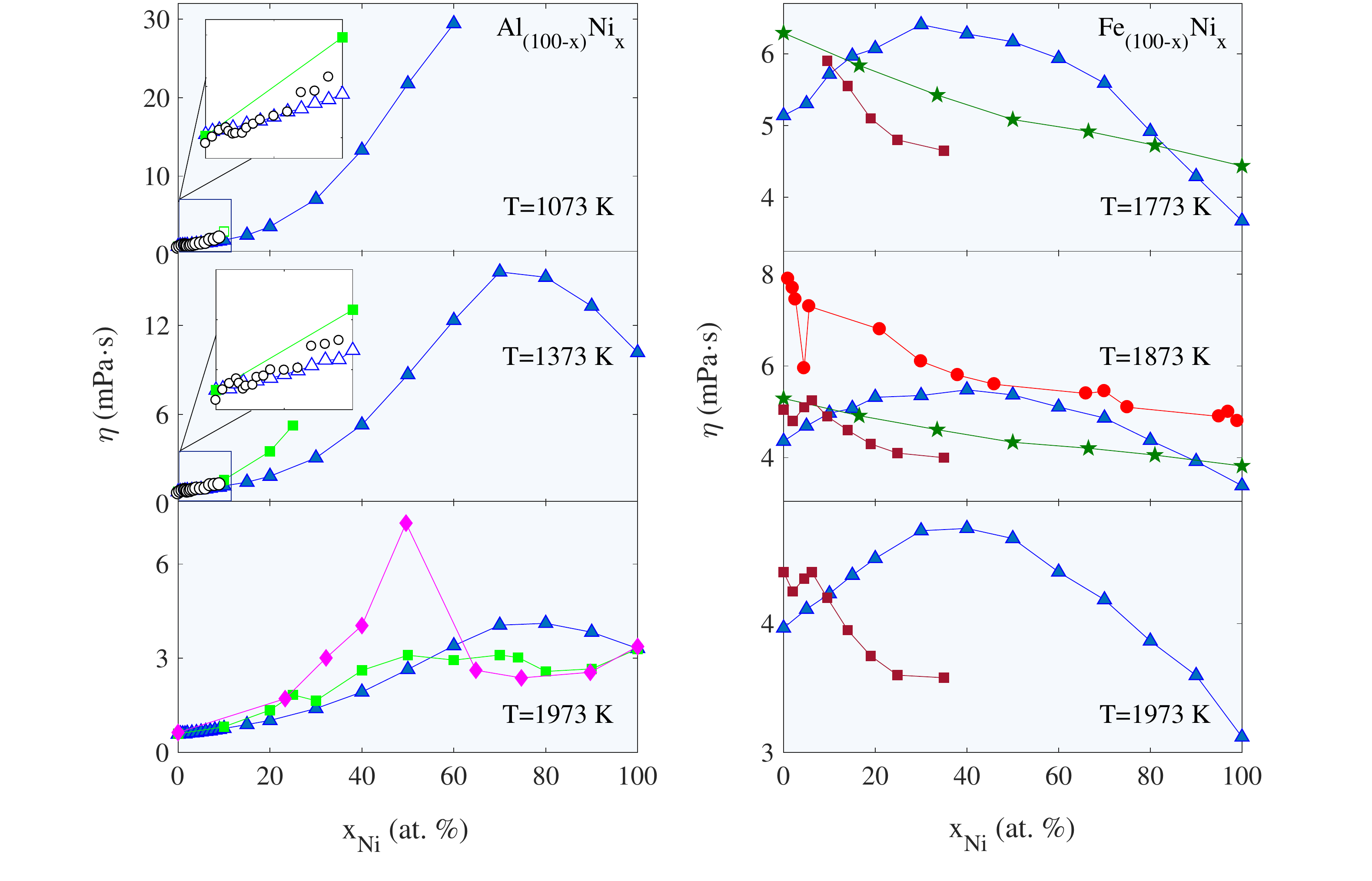}
\caption{Concentration dependences of shear viscosity of aluminum-nickel (left column) and iron-nickel (right column) melts at different temperatures: $(\triangle~\triangle~\triangle)$ -- results of molecular dynamics simulation; $(\circ~\circ~\circ)$ -- results of the experiment on viscometry; markers $(\square~\square~\square)$, $(\diamondsuit~\diamondsuit~\diamondsuit)$, $(\bigstar~\bigstar~\bigstar)$, $(\blacksquare~\blacksquare~\blacksquare)$, $(\bullet~\bullet~\bullet)$ -- experimental data taken from Refs.~\cite{Kehr2008}, \cite{Petrushevsky1971}, \cite{Sato2005}, \cite{Krieger1971}, \cite{Baum1979}, respectively.} \label{Fig_Visc3}
\end{center}
\end{figure*}

The viscosity can be calculated from an equilibrium molecular dynamics (EMD) simulation using the Green-Kubo relation
\begin{equation}
\eta=\frac{V}{k_BT}\int_0^{\infty}\langle P_{xy}(0)P_{xy}(t)\rangle dt,
\end{equation}
or the Einstein relation
\begin{equation}
\eta=\lim_{t\rightarrow\infty}\frac{V}{2tk_BT}\Bigg\langle\int_0^tP_{xy}(\tau)d\tau\Bigg\rangle,
\end{equation}
where $V$ is the simulation box volume, $k_B$ is the Boltzmann constant, $T$ is the temperature, $P_{xy}$ is the $xy$ component of the pressure tensor, and $\langle\ldots\rangle$ denotes ensemble averaging~\cite{Hansen/McDonald}. These relations are only exact in the limit of infinite simulation time and infinite simulation box length~\cite{Tenney2010}. For equilibrium metal melts, the Green-Kubo and Einstein methods yield equivalent results.

Shear viscosity $\eta$ was calculated from the reverse non-equilibrium molecular dynamics (RNEMD) method based on linear response theory~\cite{MP1999,MP2002} which provides faster convergence than the usual numerical methods~\cite{Evans/Morriss}.
The shear viscosity connects a shear field with a flux of transverse linear momentum:
\begin{equation}
j_z(p_x)=-\eta\frac{\partial\vartheta_x}{\partial z}.
\end{equation}
Here, $j_z(p_x)$ is the momentum flux and $\partial\vartheta_x/\partial z$ is a gradient of $x$-component of the fluid velocity with respect to $z$-direction. It is also denoted as the shear rate. In this method, momentum swaps were conducted between the middle and bottom bin of the simulation box, and the velocity gradient generated as a result of these momentum swaps was measured. The viscosity was then calculated as the ratio of the total flux transferred and the velocity gradient as followed
\begin{equation}
\eta=-\frac{p_x}{2tL_xL_y\langle \partial\vartheta_x/\partial z\rangle}.
\end{equation}
Here, $L_x$ and $L_y$ are the lengths of the simulation box in the $x$ and $y$ directions, respectively. The factor $2$ arises because of the periodicity of the system, and $t$ is the duration of the simulation.
The flow in the liquid is created via the \textit{fix viscosity} command in LAMMPS. The cell is divided into $20$ bins in $z$-direction, and the average velocity of group of atoms in each layer is calculated. The momentum was exchanged every $10$ time-steps. The MD trajectory lengths that are used to produce the velocity profiles are $20$~ns. The first $100$~ps are neglected due to the flow establishment.

Fig.~\ref{Fig_Visc3} shows the simulation results of the concentration dependences of the shear viscosity coefficient for aluminum-nickel (left column) and iron-nickel (right column) melts in comparison with the experimental data. The values of the experimental shear viscosity $\eta$ were obtained as $\eta=\nu\cdot\rho$, where $\nu$ is the kinematic viscosity, which is measured directly in the experiment on viscometry. The experimental values of the density $\rho$ for the $Al_{(100-x)}Ni_x$ and $Fe_{(100-x)}Ni_x$ systems were taken from the Ref.~\cite{Plevachuk2007} and \cite{Kobatake2013}, respectively. It can be seen from the figure that the simulation results are in good agreement with the results of our experiment on viscometry for aluminum-nickel melts at all the considered temperatures and for the entire studied range of concentration values. Experimental data from Ref.~\cite{Kehr2008} also show satisfactory agreement with the results of our study, while data from Ref.~\cite{Petrushevsky1971} show poor agreement. As can be seen from the graphs for the concentration dependences of $\eta(x)$ for iron-nickel melts (see the right-hand column of Fig.~\ref{Fig_Visc3}), the data of different experimental groups differ significantly. Namely, these differences amount to $1.5$ and more times. As a consequence, it is not possible to determine a general trend in the viscosity $\eta(x)$ as a function of the concentration $x$ for the $Fe_{(100-x)}Ni_x$ melts.

\begin{figure*}
\begin{center}
\includegraphics[height=7.7cm, angle=0]{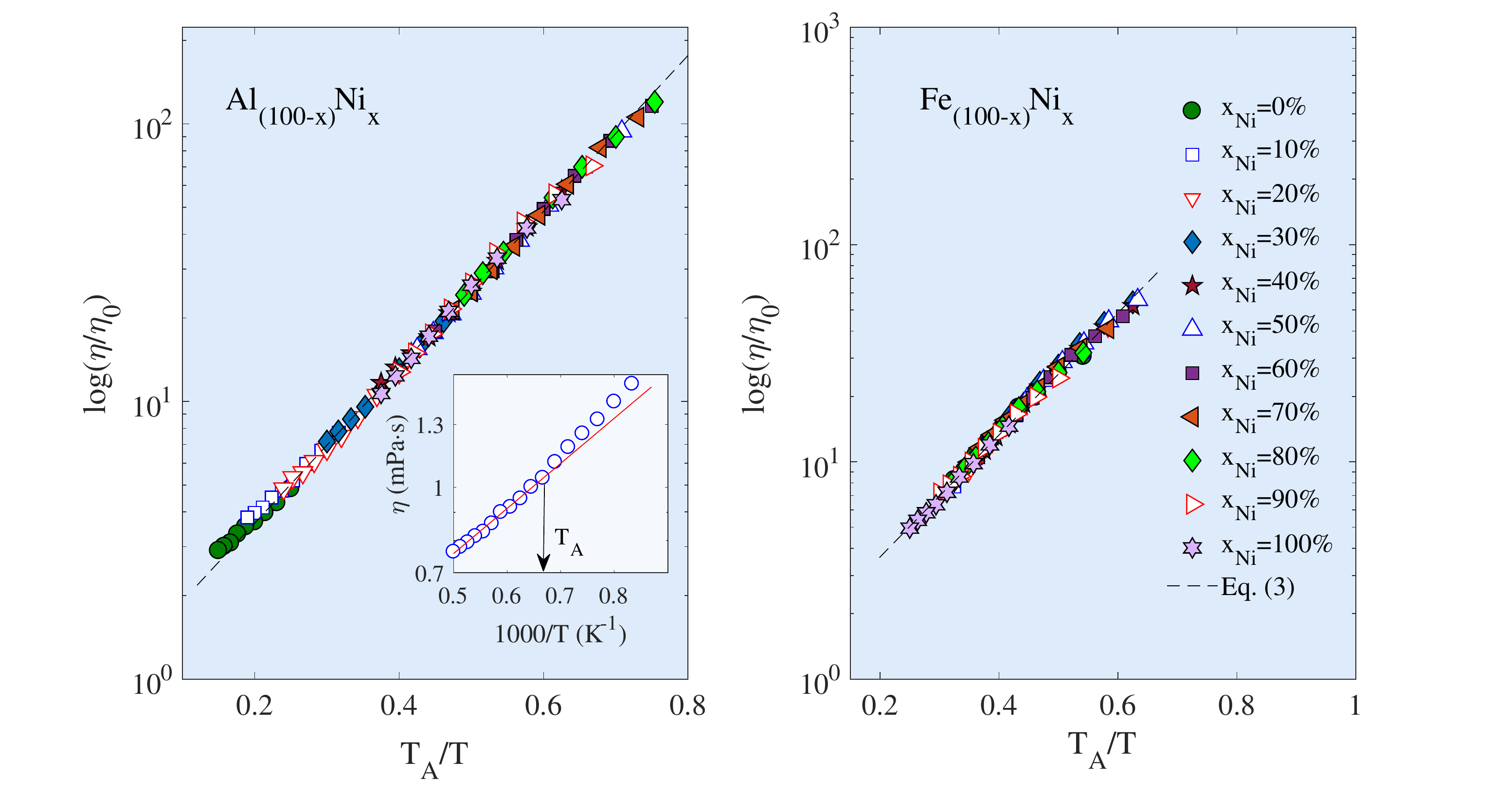}
\caption{(Color online) Temperature dependencies of the reduced viscosity for $Al_{(100-x)}Ni_x$ (left column) and $Fe_{(100-x)}Ni_x$ melts (right column) at various concentrations of nickel atoms: markers represent the results of molecular dynamics simulation; solid line -- universal viscosity curve calculated by the formula (\ref{Eq03}). \textit{Insert}: Temperature dependence of the logarithm of viscosity, showing the deviation from Arrhenius's law below the temperature $T_A$.} \label{Fig_02}
	\end{center}
\end{figure*}
In Ref.~\cite{Blodgett2015} it was shown that the temperature dependence of viscosity for a number of metal melts can be described by a universal relationship:
\begin{equation}
\eta=\eta_0\exp(E/k_BT),
\label{Eq03}
\end{equation}
\begin{equation}
E=E_{\infty}+k_BT_A(bT_r)^z\Theta(T_A-T).
\end{equation}
Here, $\eta_0$ is the pre-exponential factor, which formally corresponds to the value of the viscosity coefficient at $T\rightarrow \infty $; $E$ is the height (energy) of the activation barrier of the viscous process; $T_A$ is the Arrhenius temperature -- the temperature at which a deviation from the Arrhenius law begins to be observed in the temperature dependence of viscosity; $ \Theta(x)$ is the Heaviside function and $T_r=(T_A-T)/T_A$ is the reduced temperature.
It should be noted that there are many previously proposed expressions for the universal viscosity model such as (i) the commonly used Vogel-Fulcher-Tammann (VFT) equation \cite{Rault2000}, (ii) the recently proposed Mauro-Yue-Elliston-Gupta-Allan (MYEGA) equation \cite{Mauro2009}, (iii) the relation derived within the Cohen-Grest free volume model (CG) \cite{Cohen1979}, (iv) the avoided critical point theory (KKZNT) \cite{Kivelson1995}, (v) the cooperative shear model (DHTDSJ) \cite{Demetriou2006}, and (vi) the parabolic kinetically constrained model (EJCG) \cite{Elmatad2009}.

In Fig.~\ref{Fig_02} shows the rescaled temperature dependences of viscosity for aluminum-nickel (left column) and iron-nickel (right column) melts at various concentrations of nickel atoms. As can be seen from the figure, all temperature dependences of $\eta(T)$ are well described by the universal viscosity model \cite{Blodgett2015}. The parameters values of the model were $ E_{\infty}=6.47T_A$, $b=4.536$ and $z=2.89$. The procedure for determining the Arrhenius temperature $T_A$ is presented in the inset to the figure. Note that the Arrhenius temperature $T_A$ is associated with the deviation in the temperature dependence of viscosity from the Arrhenius behavior.

\begin{figure*}
\begin{center}
\includegraphics[height=7.7cm, angle=0]{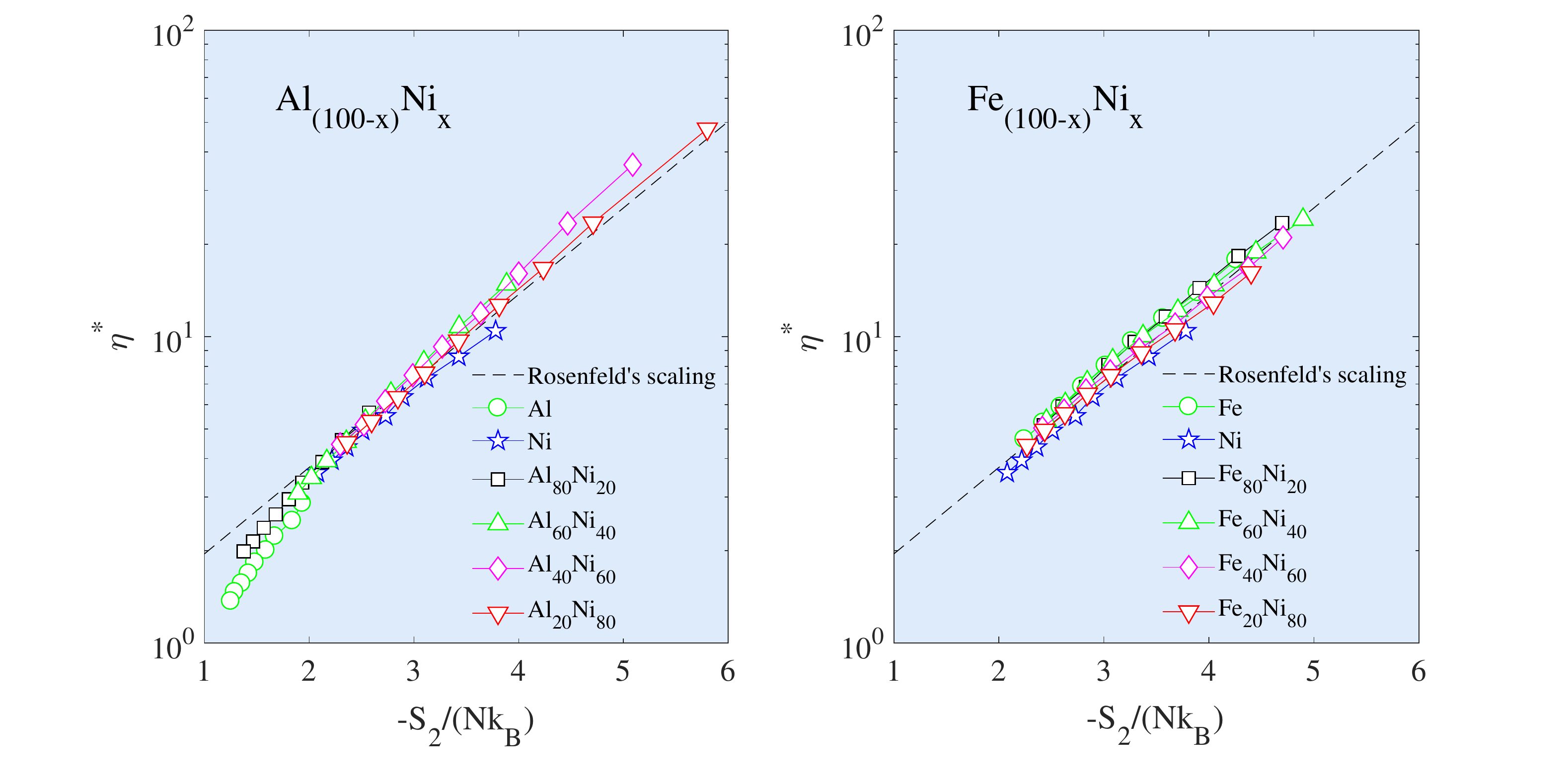}
\caption{(Color online) Reduced viscosity $\eta^*$ as a function of configurational entropy for binary nickel-containing metal melts: markers -- the results of molecular dynamics simulation; the dotted line is the Rosenfeld's scale ratio for viscosity with parameters $B=1.02$ and $\beta=0.65$.} \label{Fig_RedVisc}
\end{center}
\end{figure*}

Let us check the quasi-universality of the viscosity of binary melts depending on the composition within the framework of the Rosenfeld's scale ratio \cite{Rosenfeld1999}:
\begin{equation}
\eta^*=\eta\frac{\rho^{-2/3}}{(mk_BT)^{1/2}}=B\exp(-\beta S_{ex}),
\label{Eq_Rosen}
\end{equation}
where $\rho$ is the numerical density of the system, $m$ is the mass of atoms, $B$ and $\beta$ are dimensional coefficients, which for model systems take the values $0.2$ and $0.8$, respectively.
Here, $S_{ex}$ is the excess entropy which is defined by subtracting the ideal gas contribution $S_{id}$ from the system's entropy $S$ at the same density $\rho$ and temperature $T$, i.e.,
\begin{equation}
S_{ex}(\rho,T)\equiv S(\rho,T)-S_{id}(\rho,T).
\end{equation}
The excess entropy $S_{ex}$ is a negative quantity since the liquid is more ordered than the ideal gas.
The $S_{ex}$ in the expression (\ref{Eq_Rosen}) was replaced by the pair correlation (configurational) entropy $S_2$:
\begin{equation}
S_2=-2\pi\rho\sum_{i,j}^Nx_ix_j\int_0^{\infty}\bigg\{ g_{ij}(r)\ln[g_{ij}(r)]-[g_{ij}(r)-1]\bigg\}r^2dr,
\label{Eq_S2}
\end{equation}
where $g_{ij}(r)$ are the partial components of the radial distribution functions of atoms; the subscripts $i$ and $j$ denote the components (types of atoms) of the binary melt.
In a number of recent works \cite{Khusnutdinoff2016, Khusnutdinoff2018a, Khusnutdinoff2018b, Khusnutdinoff2011} it was shown that in the case of monatomic liquids, the two-particle contribution to the excess entropy can be $85\sim 95\%$ for a fairly wide range of densities. At the same time, for metallic melts, where contributions from many-particle interactions prevail, it is obvious that such an approximation can lead to noticeable deviations from the quasi-universal behavior of the viscosity \cite{Khusnutdinoff2018a, Khusnutdinoff2018b}. The significance of the equations obtained by Rosenfeld, in particular the equation (\ref{Eq_Rosen}), lies in providing a possible quantitative relationship between the transport characteristics (viscosity, diffusion, and thermal conductivity) and the structural features of the many-particle disordered system \cite{Dyre2018}.

It should be noted that the excess-entropy scaling has some limitations~\cite{Fomin2010,Vasisht2014,Higuchi2018}.
Liquids with anomalies like water or silica (tetrahedrally coordinated  liquids) which have, e.g., a diffusion constant which increases instead of decreases upon isothermal compression or which expand upon freezing, usually disobey the excess-entropy scaling in the ranges of the phase diagram where the anomalies appear.
It should also be noted that the Rosenfeld's and Dzugutov's scaling laws describe pure liquid metals equally well, as shown by Li et al~\cite{Li2005}.

Fig.~\ref{Fig_RedVisc} shows the reduced viscosity $\eta^*$ as a function of the configurational entropy $S_2$ for binary nickel-containing metal melts. Markers represent the molecular dynamics results, the dotted line -- the Rosenfeld's scale relation for the viscosity [equation (\ref{Eq_Rosen}) with the parameters $B=1.02$ and $\beta=0.65$]. As can be seen from the figure, for aluminum-nickel melts with a change in composition, no universal features in the behavior of viscosity are observed, while in the case of iron-nickel melts the viscosity demonstrates some general, quasi-universal features. One of the possible explanations for this discrepancy in viscosity for $Al-Ni$ melts with a change in composition may be due to the difference in the masses of aluminum and nickel atoms. For example, the component mass ratios for the considered binary metallic melts are $m_{Al}/m_{Ni}\approx 0.46$ and $ m_{Fe}/m_{Ni} \approx 0.95$.
In addition, the chemical features of the components under consideration are also different: iron and nickel belong to ferromagnetic elements with magnetic moments $\mu=2.2\mu_B$ and $\mu=0.64\mu_B$, respectively. At the same time, aluminum is a paramagnetic metal. And, as a consequence, differences will be observed in the features of the many-particle interaction of aluminum-nickel and iron-nickel melts.

Thus, we conclude that for a correct description of the viscosity of binary/multi\-component metal melts within the framework of the entropy models, it is necessary to take into account the total excess entropy $S_{ex}$, and not only its pair contribution $S_2$.

\section{Conclusions}
Large-scale molecular dynamics studies of the viscosity of aluminum-nickel and iron-nickel metal melts have been carried out. It is shown that the temperature dependencies of the viscosity of nickel-containing binary metal melts are described by a universal viscosity model. Based on the analysis of the viscosity using the Rosenfeld's scale transformations, we conclude that the use of the pair correlation entropy $S_2$ as an approximation for the excess entropy $S_{ex}$ is insufficient to correctly describe the viscosity of binary/multicomponent metal melts in the framework of entropy models.

\section{Acknowledgments}
We are grateful to Russian Science Foundation (project No. 19-12-00022). The molecular dynamic simulations were performed by using the computational cluster of Kazan Federal University and the computational facilities of Joint Supercomputer Center of Russian Academy of Sciences.

\end{document}